\documentclass[prb,superscriptaddress,twocolumn,floatfix]{revtex4} 

\usepackage{array}
\usepackage{wrapfig}

\usepackage{graphicx}
\usepackage{dcolumn}
\usepackage{bm}
\usepackage{amsmath}
\usepackage{amssymb}
\usepackage{graphics}
\usepackage{epsfig}
\usepackage{natbib}

\begin{document}


\title{Superfluid density and superconducting gaps of RbFe$_{2}$As$_2$ as a function of hydrostatic pressure}

\author{Z.~Shermadini}
\email[Corresponding author:~]{zurab.shermadini@psi.ch}
\affiliation{Laboratory for Muon Spin Spectroscopy, Paul Scherrer
Institute, CH-5232 Villigen PSI, Switzerland}
\author{H.~Luetkens}
\affiliation{Laboratory for Muon Spin Spectroscopy, Paul Scherrer
Institute, CH-5232 Villigen PSI, Switzerland}
\author{A.~Maisuradze}
\affiliation{Laboratory for Muon Spin Spectroscopy, Paul Scherrer
Institute, CH-5232 Villigen PSI, Switzerland}
\affiliation{Physik-Institut der Universit\"at Z\"urich,
Winterthurerstrasse 190, CH-8057 Z\"urich, Switzerland}
\author{R.~Khasanov}
 \affiliation{Laboratory for Muon Spin Spectroscopy, Paul Scherrer
Institute, CH-5232 Villigen PSI, Switzerland}
\author{Z.~Bukowski}
\affiliation{Laboratory for Solid State Physics, ETH Z\"urich,
CH-8093 Z\"urich, Switzerland}
\affiliation{Institute of Low
Temperature and Structure Research, Polish Academy of Sciences,
50-422 Wroclaw, Poland}
\author{H.-H.~Klauss}
\affiliation{Institut f\"ur Festk\"orperphysik, TU Dresden, D--01069
Dresden, Germany}
\author{A.~Amato}
\affiliation{Laboratory for Muon Spin Spectroscopy, Paul Scherrer
Institute, CH-5232 Villigen PSI, Switzerland}

\begin{abstract}
The superfluid density and superconducting gaps of superconducting
RbFe$_{2}$As$_2$ have been determined as a function of temperature,
magnetic field and hydrostatic pressure by susceptibility and
muon-spin spectroscopy ($\mu$SR) measurements. From the data,
fundamental microscopic parameters of the superconducting state like
the London penetration depth $\lambda$, the gap values $\Delta$, the
upper critical field $B_{c2}$, and the Ginzburg-Landau parameter
$\kappa$ have been obtained.  In accordance with earlier
measurements the ratio of the superfluid density $n_s\propto
\lambda^{-2}$ to the superconducting transition temperature
$T_c=2.52(2)$~K at ambient pressure is found to be much larger in
the strongly hole-overdoped RbFe$_{2}$As$_2$ than in high-$T_c$
Fe-based and other unconventional superconductors. As a function of
pressure $T_c$ strongly decreases with a rate of d$T_{\rm
c}$/d$p$~=~-1.32~K~GPa$^{-1}$, i.e. it is reduced by 52\% at
$p=1$~GPa. The temperature dependence of $n_s$ is best described by
a two gap \emph{s}-wave model with both superconducting gaps being
decreased by hydrostatic pressure until smaller gap completely
disappears at $p=1$~GPa.

\end{abstract}

\maketitle

\section{Introduction}

The antiferromagnetic BaFe$_{2}$As$_{2}$ is a mother compound of the
``122"  family of the iron-based arsenide unconventional
superconductors. It crystallizes in the tetragonal
ThCr$_{2}$Si$_{2}$ crystal structure with the space group of
I4/mmm.\cite{Pfisterer} High temperature superconductivity can be
induced in this system either by isovalent
substitution\cite{ShuaiJiang, Kasahara, CaoWang, Hashimoto, Qi, Ye},
chemical charge carrier doping\cite{Rotter3, Kamihara} or
hydrostatic pressure\cite{Takahashi, Mizuguchi, Zocco, Hamlin,
Okada} highlighting the large flexibility of the recently discovered
iron-based superconductors to tune its electronic properties in
general. Highest superconducting transition temperatures are
achieved by hole doping in Ba$_{1-x}$A$_{x}$Fe$_{2}$As$_{2}$ upon
substituting Ba by A~=~K\cite{Rotter3} or Rb\cite{Bukowski2,
Bukowski, Guguchia} with a gradual transition from the magnetically
ordered ground state\cite{Rotter2} towards a superconducting state
as a function of $x$. The optimal superconducting transition
temperature up to $T_{\rm c}$~$\simeq$~38~K is reached for
$x$~$\simeq$~0.4 in both cases\cite{Rotter3, Guguchia}. Different
hole and electron bands are crossing the Fermi level in case of the
optimal doped system as revealed by angle-resolved photoemission
spectroscopy.\cite{Zabolotnyy, Evtushinsky, Khasanov} The topology
of the Fermi surface is basically dominated by big hole bands around
the $\Gamma$ point and small electron bands at the $M$ point
presenting favorable nesting conditions possibly related to the
magnetic instabilities. This favorable topology is also thought to
lead to interband processes playing an important role for the
superconducting state.\cite{Ding} Indication of multi-gap
superconductivity in the hole-doped system
Ba$_{1-x}$Rb$_x$Fe$_2$As$_2$ was confirmed by our recent $\mu$SR
measurements\cite{Guguchia}. Further hole doping reduces $T_{\rm c}$
down to 3.5~K and 2.6~K for A~=~K and Rb,\cite{Sasmal} for $x$~=~1
and shifts the electron bands to the unoccupied side.\cite{Sato}
Therefore, one expects an absence of nesting conditions and hence of
magnetic order, as confirmed for example by our $\mu$SR zero field
(ZF) measurements.\cite{Shermadini} Another consequence is a strong
decrease of the interband processes which might be related to the
collapse of $T_{\rm c}$. Furthermore, it was found that the
superfluid density for the end compound of the series,
RbFe$_2$As$_2$, is much larger than in other Fe-based and
unconventional superconductors\cite{Guguchia}. This might signal a
more conventional nature of the superconducting ground state. A
further microscopic characterization of this compound seems highly
mandatory since it may, in comparison with the optimally doped
compounds from the same series, provide valuable information on the
origin of high-$T_c$ superconductivity in Fe-based materials.
Therefore, we extended our earlier $\mu$SR studies on
RbFe$_{2}$As$_{2}$ by performing measurements under hydrostatic
pressure. Due to its comparatively low upper critical field $B_{\rm
c2}$~=~2.6(1)~T and its reduced $T_{\rm c}$~=~2.52(2)~K, this system
allows to study a large section of the $B-T-p$ phase diagram. As
exemplified by a number of recent studies, the $\mu$SR technique is
very well suited to investigate iron-based superconductors (see, for
example, Ref.~\onlinecite{Amato}) and to provide quantitative
measures of several microscopic parameters of the superconducting
ground state. Here we report the determination of $T_c$, the London
penetration depth $\lambda$ and hence the superfluid density
$n_s\propto \lambda^{-2}$, the superconducting gap values $\Delta$,
the upper critical field $B_{c2}$, and the Ginzburg-Landau parameter
$\kappa$ as a function of hydrostatic pressure. From our study it
follows that the superconducting state in RbFe$_{2}$As$_{2}$ is
indeed conventional and that hydrostatic pressure pushes it even
further to the classical regime. Therefore a comparison of the
electronic properties of RbFe$_{2}$As$_2$ under pressure with
optimally doped members of the same family may provide essential
information on the origin of the high-$T_c$ phenomena in Fe-based
superconductors.

\section{Experimental Details}

The polycrystalline sample of RbFe$_2$As$_2$, used for the present
experiment, was synthesized in a two steps procedure at the
Laboratory for Solid State Physics of the
ETH-Z\"urich.\cite{Bukowski} First, RbAs and Fe$_2$As were prepared
from pure elements in evacuated and sealed silica tubes. Then,
appropriate amounts of RbAs and Fe$_2$As were mixed, pressed into
8~mm diameter pellets and annealed at 650\,$^\circ$C for several
days in evacuated and sealed silica ampoules. The quality has been
tested by x-ray diffraction and confirmed by our previous $\mu$SR
studies at ambient pressure\cite{Shermadini} that the sample is free
from the magnetic ordering. Then, these cylindrical shaped
synthesized pellets with total height of 10~mm were loaded into the
CuBe pressure cell using Daphne oil as a pressure transfer medium.
AC susceptibility measurements were performed with a conventional
lock-in amplifier at 0, 0.27, 0.46, 0.68  and 0.98 GPa pressures in
a temperature interval of 1.4-10~K, using the same pressure cell as
for $\mu$SR. Additional, magnetization data were obtained for
pressures up to 5.4~GPa on a commercial \textit{Quantum Design} 7~T
Magnetic Property Measurement System XL SQUID Magnetometer using a
home-made diamond anvil cell at temperatures between 1.8~K and 10~K.
Small lead (Pb) probes were used for pressure determination
utilizing the pressure dependence of $T_{c,Pb}$.\cite{Eiling}

The $\mu$SR measurements under pressure were performed at the
$\mu$E1 beamline of the Paul Scherrer Institute (Villigen,
Switzerland), using the GPD instrument equipped with an Oxford
sorption pumped $^{3}$He cryostat at temperatures down to 0.27~K.
The pressure cells used during the $\mu$SR measurement and
experimental setup are described elsewhere.\cite{Maisuradze,
Andreica} Data were collected at magnetic fields up to 0.25~T and
for the three pressures 0.2, 0.6 and 1.0~GPa. Zero pressure data are
taken from our early experiments.\cite{Shermadini} High-energy muons
with a momentum of 105~MeV\,c$^{-1}$ were used in order to penetrate
the pressure cell walls. The typical statistics for both forward and
backward detectors were 7~millions. The fraction of the muons
stopping in the sample was about 45\% while the remaining fraction
stops in the pressure cell. Three different transverse (TF) as well
as zero field (ZF) $\mu$SR measurements were done for each pressure
point. The temperature dependence of the superfluid density were
analyzed with the fitting package musrfit developed by A. Suter and
B. Wojek.\cite{Suter}

\section{Results and Discussion}

\begin{figure}[lt]
\center{\includegraphics[width=0.85\columnwidth,angle=0,clip]{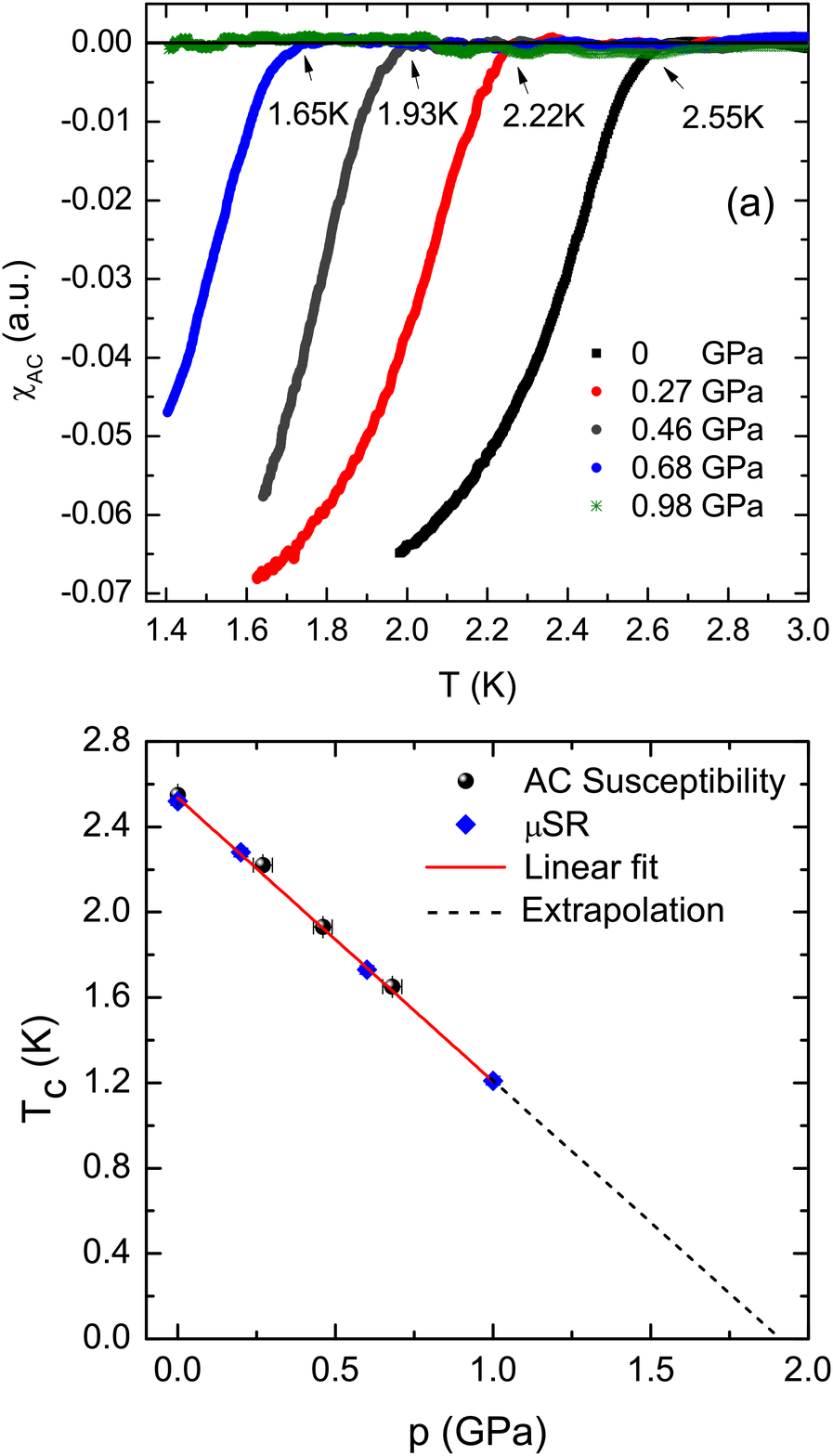}}
    \caption[]{(a)~AC susceptibility measurements up to 1.0 GPa obtained with the same CuBe pressure cell as the one used for the $\mu$SR experiments.
    (b) Pressure dependence of $T_{\rm c}$. The red solid line corresponds to a linear fit, and the
    dashed black line is an extrapolation up to 1.92 GPa.}
    \label{TcP}
\end{figure}
%

 Figure.~\ref{TcP} exibits the AC susceptibility data for pressure up
to 1.0~GPa together with the pressure dependence of $T_c$. At
ambient pressure the superconducting transition temperature of
RbFe$_{2}$As$_{2}$ is $T_{\rm c}$~=~2.55(2)K and it decreases
linearly upon increasing the pressure with a rather large value of
d$T_{\rm c}$/d$p$~=~-1.32~K~GPa$^{-1}$. A linear extrapolation of
the data suggests that superconductivity could be completely
suppressed by a pressure of approximately 2~GPa. In most
superconductors, $T_{c}$ is found to decrease under pressure, with
some exceptions as in the cuprate oxides or some iron-based systems,
which exhibit a remarkable increase.\cite{Wu, Gao} We note that a
similarly large negative slope was found in another multi-gap
superconductor MgB$_{2}$ where $T_{\rm c}$ decreases under pressure
at a rate of d$T_{\rm c}$/d$p\simeq$ - 1.11~K~GPa$^{-1}$ (see ref.
\onlinecite{Deemyad, Schilling}).

\begin{figure*}[t]
\center{\includegraphics[width=2.0\columnwidth,angle=0,clip]{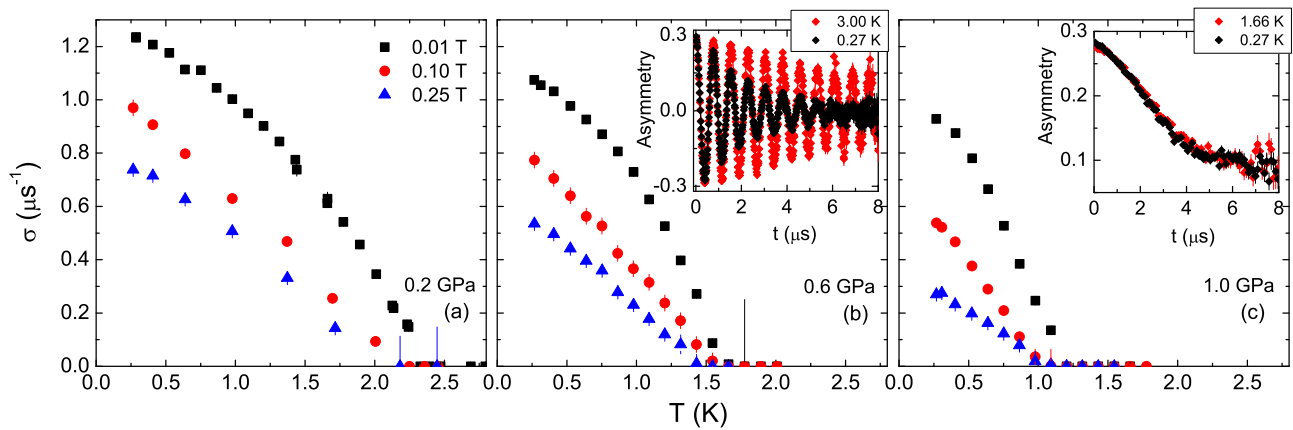}}
    \caption[]{Temperature dependence of $\sigma_{\rm s}(T)$ obtained for RbFe$_{2}$As$_{2}$ at different pressures.
    For each pressure, measurements  at different fields were performed.
    The insert on the panel (b) corresponds to TF, and on the panel (c) - ZF $\mu$SR time spectra above and below $T_{\rm c}$ at 1.0~GPa pressure.}
\label{Sigma}
\end{figure*}
The time evolution of the $\mu$SR signal is best described by the
polarization function:\cite{Maisuradze}
%
%
%
%
\begin{eqnarray}
\begin{split}
A_{0}P(t)=A_{\rm s}\exp[-\frac{(\sigma_{\rm s}^{2}+\sigma_{\rm
n}^{2})}{2}t^2]\cos(\gamma_{\mu}B_{\rm int}t+\varphi)+ \nonumber
\end{split}
\\
+A_{\rm pc}\exp[-\frac{\sigma_{\rm pc}^2}{2}t^2]\int~P(B_{\rm
pc})\cos(\gamma_{\mu}B_{\rm pc}t+\varphi)dB_{\rm pc}~,\label{eq:TF}
\end{eqnarray}
\normalsize
where the first term corresponds to the signal arising from the
sample with the corresponding parameters: $A_{\rm s}$: initial
asymmetry; $\sigma_{\rm sc}$: second moment of the magnetic field
distribution due to the flux line latice (FLL) in the mixed state;
$\sigma_{\rm n}$: depolarization rate due to the nuclear moments;
and $B_{\rm int}$: internal magnetic field at the muon site. The
second term on the right-hand side of Eq.~\ref{eq:TF} represents the
pressure cell contribution. $A_{\rm pc}$ denotes the initial
asymmetry; $\sigma_{\rm pc}$~=~0.27~$\mu$s$^{-1}$: field and
temperature independent Gaussian relaxation rate of the CuBe
pressure-cell material; and $B_{\rm pc}$ is magnetic field sensed by
the muons stopped in the pressure cell. $\gamma_{\mu}=135.5342
\times 2\pi$~MHz T$^{\rm -1}$ is the muon gyromagnetic ratio and
$\varphi$ is the initial phase of the muon spin polarization.
$P(B_{\rm pc})$ is a distribution of magnetic fields sensed by the
muons in the pressure cell which is calculated using Eq.~(A4) of
reference~\onlinecite{Maisuradze}. Therefore the integral is invoked
to describe the sum of the applied field and the field induced by
the sample in a diamagnetic state.

For each pressure point zero-field (ZF) $\mu$SR measurements have
been performed to check the magnetic properties of the system. No
sign of magnetism neither static order nor slow magnetic
fluctuations has been observed up to the highest pressure
investigated in this study. This is illustrated by the ZF spectra at
1.0~GPa shown in the insert of Fig.~\ref{Sigma}\,(c).

\begin{figure}[rt]
\center{\includegraphics[width=1.0\columnwidth,angle=0,clip]{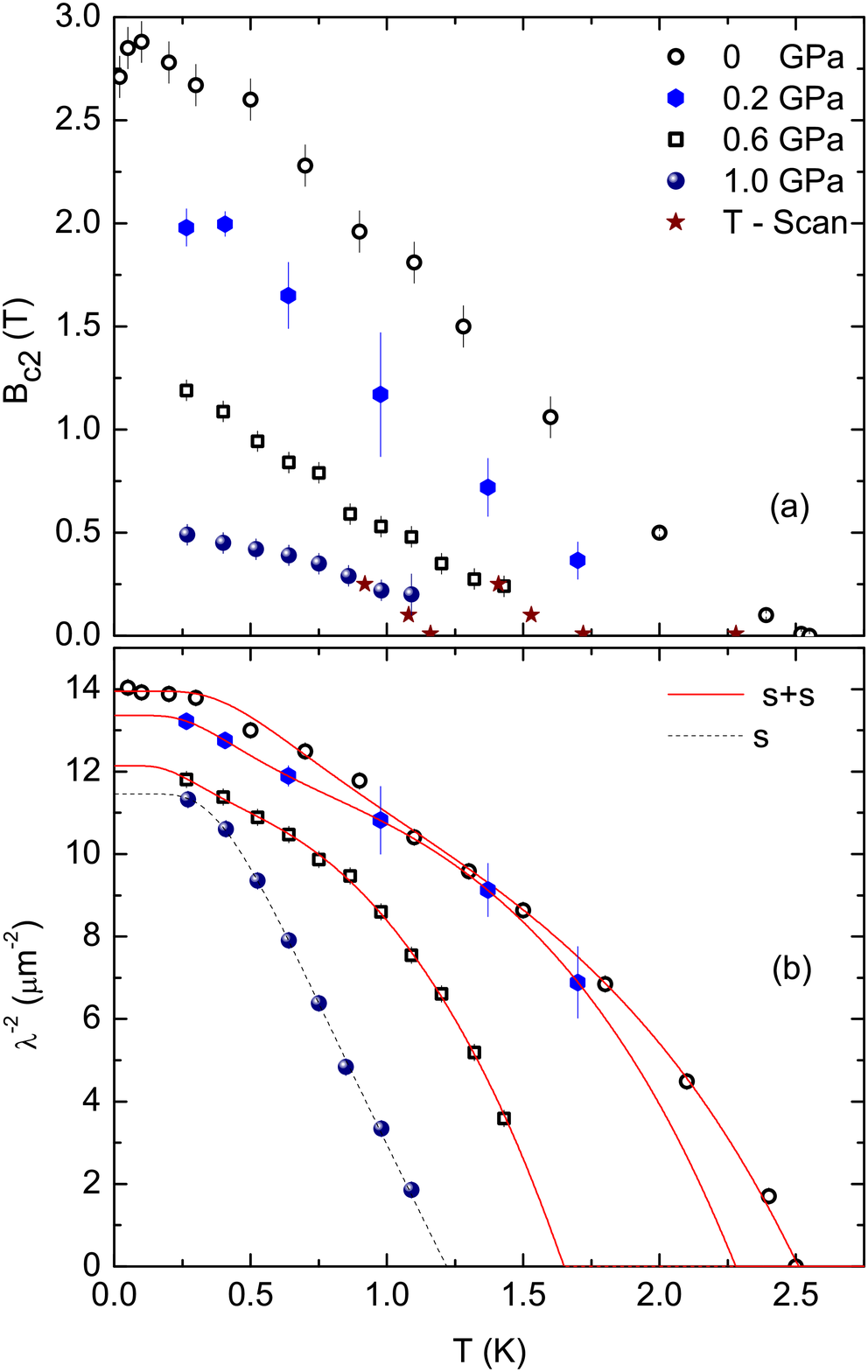}}
    \caption[]{(a)~Temperature and pressure dependence of upper the critical field $B_{\rm c2}(T)$. The stars corresponds to the values obtained
by analyzing the temperature dependence of $\sigma_{\rm s}(T)$. The
other points are obtained by fitting Eq.~\ref{BrandtFit} to the
$\sigma_{\rm s}(B_{\rm ext})|_{T,\,p\,=\,const.}$ data.
(b)~Temperature and pressure dependence of $\lambda^{-2}$(T). The
solid lines correspond to a $s+s$-wave multi-gap fit model and the
dashed one to a $s$-wave single-gap fit.} \label{Bc2L}
\end{figure}

The superconducting properties were investigated by transverse field
(TF) $\mu$SR measurements in the mixed state after field cooling
which ensures the formation of a uniform FLL. Fitting
Eq.~(\ref{eq:TF}) to the $\mu$SR time spectra, we obtain the
temperature, magnetic field and pressure dependent parameters
$\sigma_{\rm sc}(T, B, p)$ [see Fig.~\ref{Sigma}(a,b,c)]. Using the
numerical Ginzburg-Landau model developed by Brandt:\cite{Brandt}
\begin{equation}
\begin{array}{rl}
\sigma_{\rm s}\;[\mu\text{\rm s}^{-1}]=&4.83 \times 10^{4}(1-B_{\rm ext}/B_{\rm c2})\times\\
&\times[1+1.21(1-\sqrt{B_{\rm ext}/B_{\rm
c2}})^{3}]\lambda^{-2}\;[\text{nm}]\label{BrandtFit}
\end{array}
\end{equation}
one can fit the field dependent depolarization rate $\sigma_{\rm
s}(B_{\rm ext})|_{T,\,p\,=\,const.}$ and evaluate two important
parameters of the superconducting state, i.e. the London penetration
depth $\lambda$ and the upper critical field $B_{\rm c2}$ (see
Fig.~\ref{Bc2L}). This approach assumes that $\lambda$ is field
independent, as it was confirmed by our previous zero pressure
measurements.\cite{Shermadini}

\begin{figure}[lt]
\center{\includegraphics[width=1.0\columnwidth,angle=0,clip]{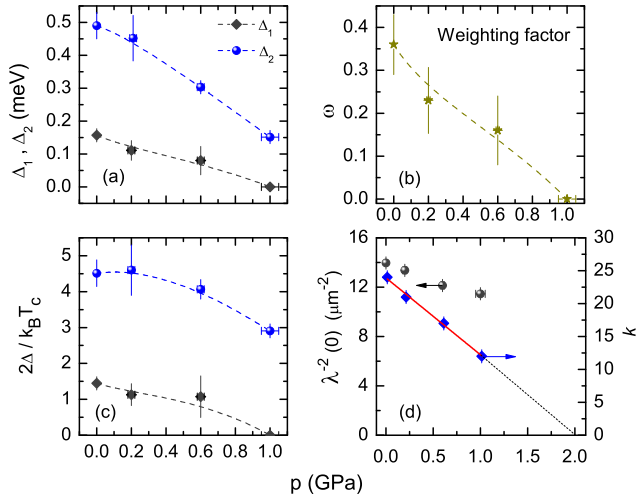}}
    \caption[]{Pressure dependence of: (a) the zero-temperature gap values
    $\Delta_{1}$(0) and $\Delta_{2}$(0), (b) the weight $\omega$ of the small gap, (c) the BCS ratios $2\Delta(0)/k_{\rm
B}T_{\rm c}$ for both gaps, and (d) the inverse square of the
penetration depth $\lambda^{-2}$(0) and Ginzburg-Landau parameter
$k$. The lines are the guides to the eyes. }. \label{Gaps}
\end{figure}

As a second step, the temperature dependence of the superconducting
carrier concentration $\rho_{\rm s}=n_{\rm s}(T)/n_{\rm
s}(0)=\lambda^{-2}(T)/\lambda^{-2}(0)$ is calculated from the
inverse square of the penetration depth. It can be fitted using the
local (London) approximation:\cite{Tinkham}

\begin{equation}
\rho_{\rm
s}=\frac{\lambda^{-2}(T)}{\lambda^{-2}(0)}=1-\frac{2}{k_{\rm
B}T}\int_\Delta^{\infty} f(\epsilon ,T)\left[ 1-f(\epsilon
,T)\right] d\epsilon~,
\end{equation}
leaving $\lambda^{-2}(0)$ and $\Delta(0)$ as free parameters. Here
$f(\epsilon ,T)~=~\left[
1+\exp\left(\sqrt{\epsilon^2+\Delta(T)^2}/k_{\rm B}T\right)\right
]^{-1}$
represents the Fermi function with the BCS spherical $s$-wave type
of $\Delta(T)$ gap.\cite{Suter} As evidenced by our previous
study\cite{Shermadini} a $s+s$ multigap model
\begin{equation}
\rho_{\rm s}=\omega~\rho_{\rm s1}+(1-\omega)~\rho_{\rm s2}
\end{equation}
where $\omega$ is the relative weighting factor for the smaller gap,
$\Delta_{1}$(0), gives a satisfactory fitting result which is also
supported by ARPES measurements where several disconnected
Fermi-surface sheets are detected for another member of 122 iron
arsenide family.\cite{Evtushinsky} The fitting results are shown in
Fig.~\ref{Bc2L}(b) and the pressure dependence of the parameters is
reported on Fig.~\ref{Gaps}. It is found that the hydrostatic
pressure only slightly reduces the superfluid density. Note that at
1.0~GPa, one does not require anymore the $s+s$ model and that a
single $s$-wave gap scenario (i.e. $\omega \simeq 0$) is sufficient
to describe the data. One may remark that upon increasing the
hydrostatic pressure the positive curvature of $B_{\rm c2}$(T) near
$T_{\rm c}$ gradually disappears and ends up with a usual BCS
temperature dependence shape at 1.0~GPa~[Fig.~\ref{Bc2L}(a)], giving
an additional indication of the disappearance of the smaller gap.
Both gaps gradually decrease and at 1.0~GPa the small gap
$\Delta_{1}$ essentially disappeared [Fig.~\ref{Gaps}(a)];
correspondingly, its weighting factor is falling from a maximum of
$\omega$~=~0.36 value to 0 [Fig.~\ref{Gaps}(b)]. The BCS ratios
$2\Delta_{1}(0)/k_{\rm B}T_{\rm c}$\,=\,1.5(1) and
$2\Delta_{2}(0)/k_{\rm B}T_{\rm c}$\,=\,4.5(1) are relatively
independent on pressure up to 0.6~GPa followed by a gradual drop for
the large gap value in the absence of the smaller gap.
Upon increasing the hydrostatic pressure from 0 to 1.0~GPa, $T_{\rm
c}$ is reduced by $\sim$~52\%, while the superfluid density $\rho
\propto \lambda^{-2}$ is decreased by $\sim$~18\%, only. In other
words, as shown in Fig.~\ref{Gaps}~(d), the superfluid density only
weakly depends on the hydrostatic pressure in contrast to the strong
dependence of $T_{\rm c}$, typical for unconventional high-$T_c$
superconductors where a proportionality of this two quantities is
usually observed (at least in under- and optimally doped compounds).
Using the pressure-dependent values of the penetration depths and
upper critical field one can determine the characteristic ratio,
known as the Ginzburg-Landau parameter, $\kappa=\lambda / \xi$,
determined at our base temperature of 0.27~K.\cite{Tinkham} $\xi$ is
a superconducting coherent length calculated from the relation
$B_{\rm c2}$~=~$\Phi_{0}$/$2\pi\xi^{2}$, where
$\Phi_{0}$~=~2.0678$\times$10$^{-15}$~Wb is the magnetic flux
quantum. A reduction of 50\% of $\kappa$ is determined at the
highest available pressure of this experiment, pointing to a clear
shift of the superconducting character of RbFe$_2$As$_2$ away from a
strong type II superconductor towards low $\kappa$ classical BCS
superconductors. Interestingly, both $T_c$ and $\kappa$ linearly
decrease with pressure and therefore we find the experimental
correlation $\kappa\propto T_c$ (see Table~\ref{table1}).
\begin{table}[b]
\caption{List of the pressure dependent parameters obtained from the
analysis of $\lambda^{-2}$(T).}
\centering
\begin{ruledtabular}\begin{tabular}{c  c  c  c  c  c  c  c}
$p$ & $T_{\rm c}$ & $\Delta_{1}$(0) & $\Delta_{2}$(0) &
$\frac{2\Delta_{1}(0)}{k_{\rm B}T_{\rm c}}$ &
$\frac{2\Delta_{2}(0)}{k_{\rm
B}T_{\rm c}}$ & $\kappa$ & $\lambda(0)$\\
(GPa)& (K) & (meV) & (meV) & & & & (nm) \\
\hline 
0.00(0) & 2.52(2) & 0.15(2) & 0.49(4) & 1.5(2) & 4.5(4) & 24(1) &
267(5) \\
0.20(1) & 2.28(1) & 0.11(3) & 0.45(7) & 1.1(3) & 4.6(7) & 21(1) &
274(5) \\
0.60(1) & 1.73(1) & 0.08(4) & 0.30(2) & 1.1(6) & 4.1(3) & 17(1) &
287(6) \\
1.00(2) & 1.21(1) & 0.00(0) & 0.15(1) & 0.0(0) & 2.9(2) & 12(1) &
295(6) \\
\end{tabular}\end{ruledtabular}
\label{table1} 
\end{table}
%

%

\begin{figure}[rt]
\center{\includegraphics[width=1.0\columnwidth,angle=0,clip]{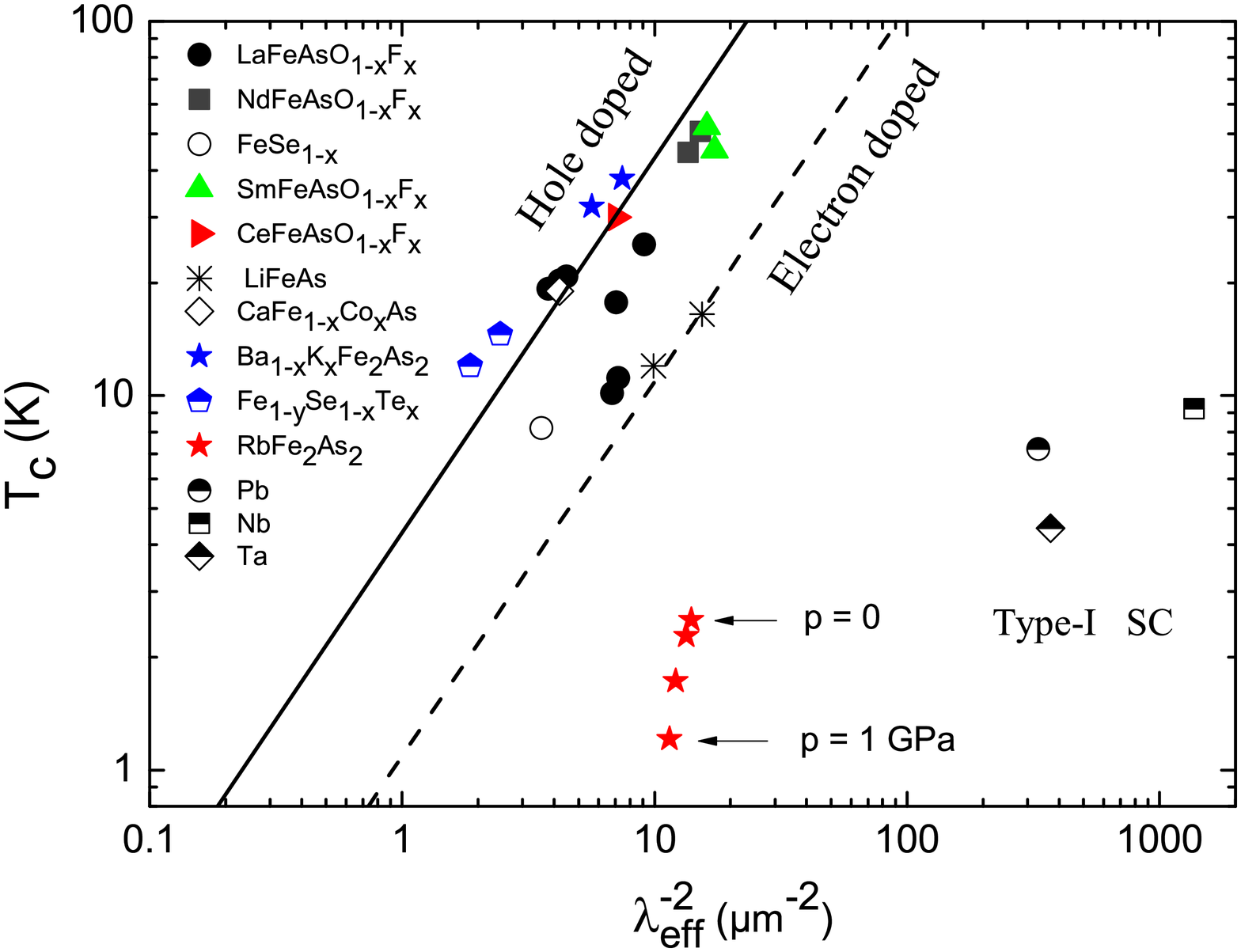}}
    \caption[]{Uemura plot for some Fe-based high temperature superconductors.
The Uemura relation observed for underdoped cuprates is shown as a
black dashed line for electron doping and as a solid line for hole
doping [\onlinecite{Shengelaya}]. The data are taken from the
following references:  LaFeAsO$_{\rm 1-x}$F$_{\rm x}$ -
[\onlinecite{Luetkens, Luetkens2, Takeshita, Carlo}]. NdFeAs$_{\rm
1-x}$F$_{\rm x}$ - [\onlinecite{Carlo, khasanov3}]. FeSe$_{\rm 1-x}$
- [\onlinecite{khasanov2, khasanov4}]. SmFeAs$_{\rm 1-x}$F$_{\rm x}$
- [\onlinecite{khasanov3, Drew}]. CeFeAs$_{\rm 1-x}$F$_{\rm x}$ -
[\onlinecite{Carlo}]. LiFeAs - [\onlinecite{Pratt}]. CaF$_{\rm
1-x}$Co$_{\rm x}$Fe$_{\rm 2}$As$_{\rm 2}$ -
[\onlinecite{Takeshita2}]. Ba$_{\rm 1-x}$K$_{\rm x}$Fe$_{\rm
2}$As$_{\rm 2}$ - [\onlinecite{Khasanov}]. Fe$_{\rm 1-y}$Se$_{\rm
1-x}$Te$_{\rm x}$ - [\onlinecite{Kim, Bendele}]. Pb, Nb and Ta -
[\onlinecite{Suter2}]. RbFe$_{\rm 2}$As$_{\rm 2}$ - obtained in this
work and in [\onlinecite{Shermadini}].} \label{Uemura2}
\end{figure}

One way to visualize and to shed light onto the nature of the
superconducting state has been presented by Uemura {\it et al.} in
Refs.~\onlinecite{Uemura} and  \onlinecite{Uemura1}. According to
the so-called ``Uemura plot" the universal linear relation between
$T_{\rm c}$ and $\sigma_{\rm s}(T \rightarrow 0)$ has been found for
high temperature superconductor cuprates. The critical temperature
appears to be proportional to the inverse square of the London
penetration depth $T_{\rm c}\propto\rho_{\rm s}\propto\lambda^{-2}$
for a large number of cuprate superconductors, but the
proportionality constant is different for hole- and electron-doped
superconductors.\cite{Shengelaya} A number of Fe-based
superconductors appear to follow the Uemura relation (see
Fig.\ref{Uemura2}). For comparison reason we include the data points
of RbFe$_2$As$_2$ to the Uemura plot. As evidenced from
Fig.~\ref{Uemura2}, various families of unconventional
superconductors including high-$T_c$ Fe-based materials are
characterized by small $\lambda^{-2}$ values (superfluid density)
compared to their $T_c$; i.e. they exhibit a dilute superfluidity.
In contrast, conventional phonon mediated superconductors like
elemental metals possess a dense superfluid, and exhibit low values
of $T_c$. RbFe$_2$As$_2$ falls in between these two extreme cases.
With increased hydrostatic pressure the critical temperature reduces
rapidly, compared to superfluid density, and the relation of $T_{\rm
c}$ to $\lambda^{-2}$ moves closer to the one characteristic for
conventional superconductors.

 In the following, we discuss the different effects of chemical and
hydrostatic pressure on the RbFe$_2$As$_2$ system. As mentioned
above, the related compound KFe$_2$As$_2$ is also superconducting
with an increased $T_c=3.8$~K compared to $T_c=2.6$~K in
RbFe$_2$As$_2$. Due to the smaller ionic radius of K$^+$ compared to
Rb$^+$ both the \emph{a}- and \emph{c}-axis parameters of
KFe$_2$As$_2$ are reduced.\cite{Sasmal, Bukowski2, Rotter, Rotter2,
Rotter3} In other words, in the up-to-now hypothetical series
Rb$_{1-y}$K$_y$Fe$_2$As$_2$ the increasing chemical pressure with
increasing $y$ should finally lead to the experimentally observed
increased $T_c$. In sharp contrast, our hydrostatic pressure
experiments on RbFe$_2$As$_2$ show a strong reduction of $T_c$ with
increasing pressures. A possible way out of the apparent discrepancy
could be a non-monotonic dependence of $T_c$ on pressure with a
reappearance of superconductivity at higher pressures as it was
recently observed in another Fe-based superconductor.\cite{GuoJing,
Liling} Therefore, we tested this hypothesis by performing further
magnetization studies under high pressure using a diamond anvil
cell. No superconducting transition was detected above 1.8~K up to
our maximum pressure of 5.4~GPa. Based on these experimental facts,
one has to conclude that external pressure is not equivalent to
chemical pressure in this particular compound. This is probably
related to the different effects of the two forms of pressure on the
local atomic structure within the FeAs tetrahedra which is known to
be one of the governing parameters determining $T_c$ in Fe-based
superconductors.\cite{Mizuguchi2}


\section{Summary and Conclusion}

 To summarize, $\mu$SR and magnetization measurements under hydrostatic pressures
(0.2,~0.6 and 1.0~GPa) were carried out on the polycrystalline
RbFe$_{2}$As$_2$ hole-overdoped iron-based superconductor. A
negative pressure effect was observed on critical temperature with a
rate of d$T_{\rm c}$/d$p$~=~-1.32~K~GPa$^{-1}$ in contrast to a
positive effect expected for an equivalency of chemical and
hydrostatic pressures. The zero temperature values of London
penetration depth $\lambda$(0), superconducting gaps $\Delta$(0),
upper critical field $B_{\rm c2}$, and Ginzburg-Landau parameter
$\kappa$~=~$\lambda / \xi$ have been evaluated from the experimental
data. The superfluid density was found to be weakly pressure
dependent, while $\kappa$ and $T_{\rm c}$ are linearly reduced by
50\% by the application of pressures up to 1~GPa. Upon increasing
the hydrostatic pressure, the system undergoes a transition from a
$s+s$-wave multi-gap superconducting state to a single $s$-wave gap
state.

Here, one can highlight three main points in favor of tendency to
the conventional BCS type of superconductors:

 (i) Upon increasing the hydrostatic pressure RbFe$_{2}$As$_2$ compound
 exhibits a gradual transition from two gap to single gap state ending up with the BCS ratio
of $2\Delta/k_{\rm B}T_{\rm c}$~=~2.9(2).

 (ii) A Strong reduction of $\kappa$ from 24 down to 12 is observed,
getting closer to the conventional BCS superconductors, and in the
limit of high pressures it extrapolates to value typical for the
type I superconductors.

 (iii) The Uemura classification scheme shows that with increased hydrostatic pressure
the critical temperature reduces more rapidly than the superfluid
density, and the relation of $T_{\rm c}$ to $\lambda^{-2}$ moves
closer to the region where low critical temperature and high
superfluid density are characteristics for conventional
superconductors.

Moreover, $n_s$ is only diminished by 18\% at $p=1$~GPa indicating
that the proportionality of $n_s$ and $T_c$ found for several
families of under and optimally doped unconventional superconductors
does not hold for RbFe$_{2}$As$_2$ either. On the other hand, these
observations are rather typical for classical low temperature BCS
superconductors.\cite{khasanov5} In addition, the temperature
dependence of $n_s$ is best described by a two gap \emph{s}-wave
model with both superconducting gaps being decreased by hydrostatic
pressure until the smaller gap completely disappears at 1~GPa.
Hence, the hydrostatic pressure appears to shift the nature of the
ground state of the hole-overdoped RbFe$_{2}$As$_2$ system to an
even more classical superconducting state. The superconducting
ground state of the hole-overdoped RbFe$_{2}$As$_2$ system appears
to be rather conventional. Since no superconducting transition was
detected above 1.8~K up to 5.4~GPa pressure, one may conclude that
external pressure is not equivalent to chemical pressure in this
particular compound. The experimental and theoretical comparisons of
the electronic properties of RbFe$_{2}$As$_2$ under pressure with
optimally doped members of the same family should therefore provide
new insight into the origin of the high-$T_c$ phenomena in Fe-based
superconductors.

\section{Acknowledgments}

  The $\mu$SR and magnetization  experiments up to 1.0~GPa were performed at the Swiss Muon Source,
Paul Scherrer Institute, Villigen, Switzerland. We acknowledge
support by the Swiss National Science Foundation, the NCCR Materials
with Novel Electronic Properties (MaNEP).

\end{document}